\documentclass[a4paper,11pt]{article}
\usepackage{pos}
\usepackage{listings}
\usepackage{enumitem}
\usepackage{setspace}
\usepackage{multicol}

\definecolor{codebg}{RGB}{240,240,240}

\lstdefinestyle{custom}{
    backgroundcolor=\color{codebg},
    basicstyle=\ttfamily\small,
    numbers=left,
    numberstyle=\tiny,
    breaklines=true,
    frame=single
}

\title{A Julia Code for Lattice QCD on GPUs}

\author[a]{Guilherme Catumba}
\author*[b]{Fernando P. Panadero}
\author[b,c]{Carlos Pena}
\author[a]{Alberto Ramos}

\affiliation[a]{Instituto de Física Corpuscular (IFIC), CSIC-Universitat de València, 46071, Valencia, Spain}
\affiliation[b]{Instituto de Física Teórica UAM-CSIC, c/ Nicolás Cabrera 13-15}
\affiliation[c]{Department of Theoretical Physics, Universidad Autónoma de Madrid, 28049 Madrid, Spain}

\emailAdd{fernando.p@csic.es}

\abstract{We present a new GPU-based open source package to perform
  Lattice simulations developed in Julia. The code currently supports
  generation of SU(2) and SU(3) (pure gauge) configurations with
  different actions and boundary conditions, and is able to perform
  measurements of flow observables (both gluonic and fermionic) as
  well as different fermionic two point functions. We will show the
  capabilities of the package, and provide information about some
  measurement codes built on top of this framework.} 

\FullConference{The 41st International Symposium on Lattice Field Theory (LATTICE2024)\\
  28 July - 3 August 2024\\
  Liverpool, UK\\}

\begin{document}
\maketitle

\section{Introduction}

Lattice simulations provide a framework for first-principles
computations in strongly coupled quantum field theories. Highly
salable codes for HPC systems and algorithmic developments play a key
role in this field, specially in the most demanding task of making
precise computations in Quantum Chromodynamics (QCD).

While most lattice simulation software has been designed for CPUs,
rapid development of Graphics Processing Units (GPUs) provides very
efficient tools, which has led to a sustained effort both to adapt the
simulation codes to support GPUs, and develop new software based
directly on them. This is the main motivation for developing our code.

Julia \cite{Julia-2017} is a high-level programing languaje that
provides several atractive features. The main one is just-in-time
(JIT) compilation, resulting in a performance close to standard
compiled languages like C while maintining the code easy to both read
and develop. Multiple dispatch is also a very usefull feature for both
developing and using the package, since it allows for a high degree of
abstraction. Julia also provides a built-in package manager, which is
fundamental for the compatibility and development of any package.

\texttt{LatticeGPU.jl} \cite{LGPU} is an open source package developed
in Julia, designed to perform lattice simulations using GPUs. This
code aims to provide a good balance between the high computational
efficiency required to perform lattice simulations in reasonable time
periods, and the simplicity and speed in code development required to
implement and test new ideas.

The code does not support GPU
parallelization yet, so no interconnect between GPUs is assumed. This
still allows for some parallelization in lattice simulations, since
different sections of the Monte Carlo chain, or replica runs, can be
processed in parallel by different GPUs.

Some of the more relevant features of the code include the generation
of quenched configurations for SU(2) and SU(3) for different actions
and boundary conditions, the computation of O(a) improved Wilson
fermion propagators and contractions, and the meassurement of a number
of observables in both the gluonic and fermionic sectors. Some of
these capabilities will be described in this work, while more details
about the full content of the code will be available in an upcoming
publication. Several checks have been performed both against published
results \cite{Luscher:2010iy,Haan:1987xa, Jansen:2003ir,
  Guagnelli:2004za, Luscher:2013cpa}, other codes \cite{oqcd}, and
consistency checks available within the package itself.

\section{Code structure}

In this section, we will discuss some key features available in the
package. While not every detail will be covered, the goal is to
provide an overview of the main tools available and the overall
usability of the code.

\subsection{Lattice geometry}

The geometry of the lattice is encoded in the \texttt{SpaceParm} structure, needed for most functions in the package. A variable of this type can be defined via the contructor
\begin{center}
  \texttt{lp = SpaceParm\{D\}(iL,sub\_iL,BC,Tw\_tensor),}
\end{center}
\vspace{1.5mm}
\noindent
\begin{minipage}{0.45\textwidth}
where \texttt{D} is the number of dimensions, \texttt{iL} is the dimension of the lattice, \texttt{sub\_iL} is the dimension of the sub-blocks, \texttt{BC} defines the boundary conditions and \texttt{Tw\_tensor} is the twist-tensor \cite{tHooft:1979rtg} (which can be ommited in the constructor).

The lattice size \texttt{iL} must be a tuple of \texttt{D} integers, where the last integer will denote the time extent. The dimension of the sub-blocks \texttt{sub\_iL} must be a tuple of the same type, where each component must divide the dimension of the lattice in the same direction and will define how the different fields are parallelized inside the GPU. This parallelization is illustrated in Figure \ref{spaceparal}.
\end{minipage}\hfill
\begin{minipage}{0.5\textwidth}
  \centering
  \includegraphics[width = \linewidth]{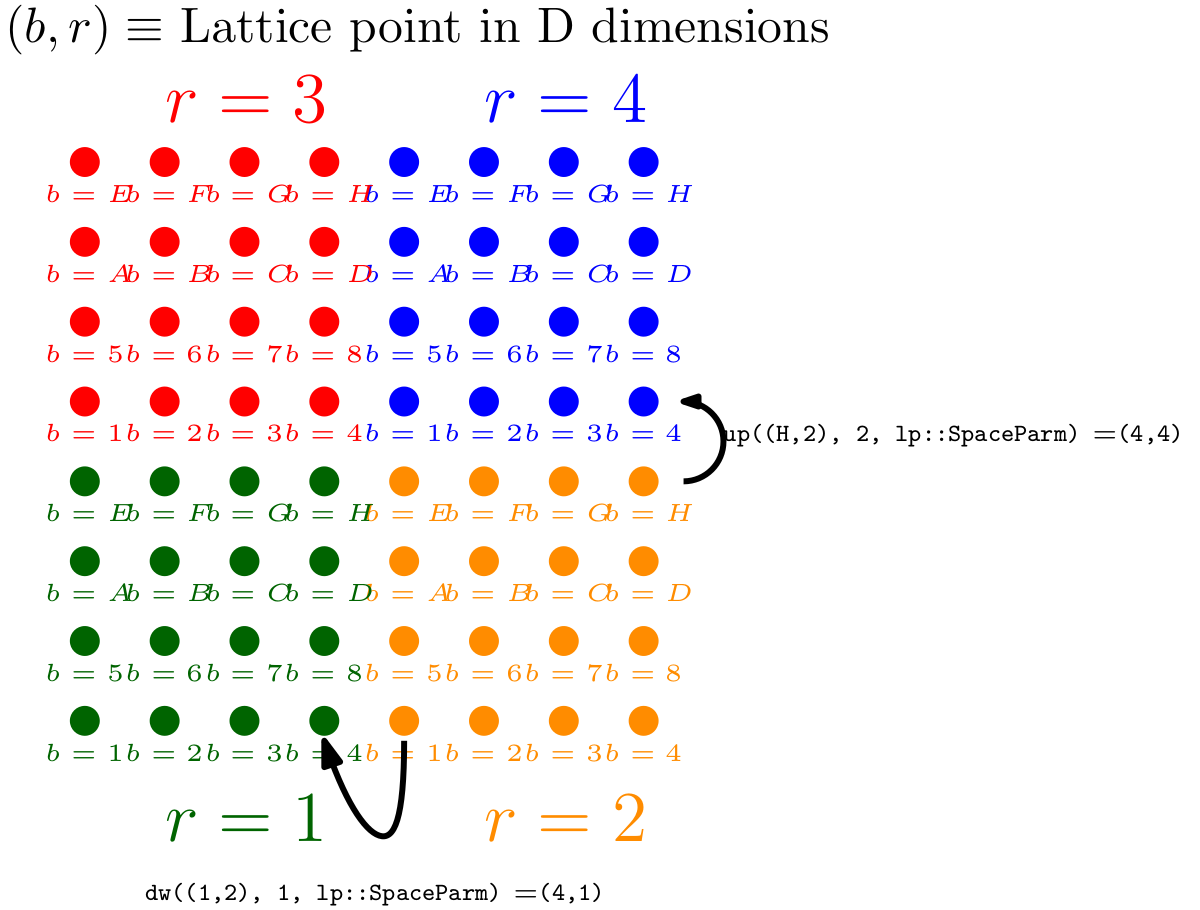}
  \captionof{figure}{An example on how the lattice geometry is parallelized in the GPU. In this case, the lattice geometry is defined by \texttt{SpaceParm\{2\}((8,8),(4,4),BC,Tw\_tensor)}. The functions \texttt{up} and \texttt{dw}, used to move with the "GPU coordinates", are also illustrated.}
  \label{spaceparal}
\end{minipage}
\vspace{1.5mm}

The available options for boundary conditions are the following:
\begin{itemize}[itemsep=-0.3pt]
  \item \texttt{BC\_PERIODIC}: Periodic boundary conditions in all space-time directions.
  \item \texttt{BC\_SF\_ORBI}: Schrödinger functional (SF) boundary conditions, orbifold construction \cite{Luscher:1992an}.
  \item \texttt{BC\_SF\_AFWB}: SF boundary conditions, Aoki-Frezzoti-Weisz choice B \cite{Aoki:1998qd}.
  \item \texttt{BC\_OPEN}: Open boundary conditions in the time direction \cite{Luscher:2011kk}.
\end{itemize}

For the case of open boundary conditions, the effective size of the time direction will be \texttt{iL[4]-1}, while for the Periodic and both cases of Schrödinger functional boundaries, the effective time extent is \texttt{iL[4]}.

\subsection{Fields and data structure}

The relevant fields for a simulation will be CUDA arrays (\texttt{CuArray}) permanently stored in the GPU. The dimension and element-type of this array will depend on the field we are working with, and in most cases the elements will be custom structures of the package.

The elements of the different arrays will map into the lattice geometry according to the parallelization defined in the \texttt{SpaceParm} structure, while additional indices can label features like the direction (vector fields) or plane indexing (tensors fields). These \texttt{CuArray} can take values of several different structures, the two most relevant cases being gauge fields and fermion fields. In the case of gauge fields, one uses \texttt{vector\_field} with values in a specific group. The two available choices are \texttt{SU3\{T\}} and \texttt{SU2\{T\}}, where \texttt{T} denotes the precision, usually \texttt{Float64}.

The way these stuctures store a group element is designed to minimize memory use: for \texttt{SU2} we use the Caley-Dickson representation ($g=(z_{1},z_{2}), |z_{1}|^{2}+|z_{2}|^{2}=1$), while for \texttt{SU3} we store the first two rows of the matrix and recompute the third row every time it is needed, using unitarity.

For the case of (pseudo-)fermion fields, the data structure is nested in the following way: \texttt{CuArray > Spinor\{NS,REP\} > REP > Complex\{T\} > T}. The type parameter \texttt{NS} will tipically be equal to four and \texttt{REP} can take the values \texttt{SU2fund\{T\}} and \texttt{SU3fund\{T\}}.

Most functions in the package require the use of auxiliary fields,
that should be allocated in the GPU beforehand. This allocation is
done in the corresponding workspace structures: the one related to the
gauge sector can be allocated via \texttt{ymws =
  YMworkspace(SU3,Float64,lp)}, while for the fermionic sector we have
\texttt{dws = DiracWorkspace(SU3fund,Float64,lp)}. In both cases,
replacing \texttt{SU3} for \texttt{SU2} is also supported. This
workspace  is designed to minimize memory allocations. While this
permanent allocation together with the non-parallelization between
GPUs imposes a bound on the available lattice volumes, the code can
comfortably run simulations on a $L/a=64$ volume with 40 GB of GPU
memory.

It is possible to use the allocated fields in intermediate steps of a
main program in order to avoid allocating too many variables, but this
practice is not recommended since these fields will be overwritten by
some routines of the package. The complete list of allocated fields in
each workspace, as well as the methods that modify them, can be be
found in the documentation.

\section{Available features}

\subsection{Monte Carlo generation}

The package supports so far the generation of quenched configurations with the Hybrid Monte Carlo algorithm. The gauge action in the bulk of the lattice can be written as

\begin{equation}
  S_{G} = \beta \sum_{x,\mu > \nu} c_{0}Tr(\mathbf{1} - P_{\mu \nu}) + c_{1}Tr(\mathbf{1} -  R_{\mu \nu}),
\end{equation}where $P_{\mu \nu}$ and $R_{\mu \nu}$ are respectively
the oriented $1\times 1$ and $1\times 2$ Wislon loops. The condition
$c_{0} + 8c_{1} = 1$ is assumed. This
action parametrizes a subset of a general O(a) improved action for the
gauge fields, such as L\"uscher-Weisz\cite{Luscher:1984xn} or
Iwasaki\cite{Iwasaki:1983iya}. The relevant structure to define the
parameters of the action is the \texttt{GaugeParam} structure, and can
be defined by \texttt{gp =
  GaugeParm\{Float64\}(GRP,beta,c0,(cG0,cG1),phi,lp.iL),} where
\texttt{GRP} is the gauge group, \texttt{beta} is the value of the
inverse gauge coupling ($2N/g_{0}^{2}$), \texttt{c0} is defined in the
action, the parameters \texttt{cG0} and \texttt{cG1} are the weights
for the action in the boundaries. The boundary values of the spatial
links for the SF boundary conditions are defined by \texttt{phi
}$=(\phi_1,\phi_2)$ via\footnote{Note that this parametrization is
  only available for \texttt{SU3}.} $U_{i} =
exp(i\phi_{1}/lp.iL[i],i\phi_{2}/lp.iL[i],-i(\phi_{1}+\phi_{2})/lp.iL[i])$.

The Hybrid Monte Carlo method is implemented in the function

\begin{equation*}
  \texttt{dh,acc = HMC!(U,intsch,lp,gp,ymws).}
\end{equation*}

The first variable is the gauge field configuration, that will be modified. The second variable defines the integration scheme for the Hamilton equations. Avaiable options are: \texttt{leapfrog(Float64,} \texttt{epsilon,number\_of\_steps)}, and the Omelyan integrators \texttt{omf2(...)} and \texttt{omf4(...)}. The last input is the Yang-Mills workspace, explained in the previous section.

The \texttt{HMC!} function will return two variables: the energy violation of the process and a boolean true if the configuration was accepted (and false if it was rejected).

\subsection{Gauge observables}

The main functions for the measurements of observables involving the Yang-Mills sector are the following

\begin{itemize}[itemsep=-0.3pt]
  \item \texttt{plaquette(U, lp::SpaceParm, gp::GaugeParm, ymws::YMworkspace)}: Returns the average value of the plaquette.
  \item \texttt{function gauge\_action(U, lp::SpaceParm, gp::GaugeParm, ymws::YMworkspace)}: Returns the value of the gauge action.
  \item \texttt{function sfcoupling(U, lp::SpaceParm, gp::GaugeParm, ymws::YMworkspace)}: Measures the Schrödinger Functional coupling $dS/d\eta$ and $d^{2}S/d\eta d\nu$ \cite{Luscher:1992an, Sint:1993un}.
  \item \texttt{function Eoft\_plaq([Eslc,] U, gp::GaugeParm, lp::SpaceParm, ymws::YMworkspace)}: Returns the value of the action density using the plaquette discretization.
  \item \texttt{function Eoft\_clover([Eslc,] U, gp::GaugeParm, lp::SpaceParm, ymws::YMworkspace)}: Returns the value of the action density using the clover discretization.
  \item \texttt{function Qtop([Qslc,] U, gp::GaugeParm, lp::SpaceParm, ymws::YMworkspace)}: Returns the value of the topological charge with the clover definition.
\end{itemize}

The first argument in the last functions can be ommited; if included, it must be a vector of size \texttt{lp.iL[4]}, and the contribution of each time-slice will be written in that vector.

The gradient flow \cite{Luscher:2010iy} has proven to be a very powerful tool for many years, specially in the Yang-Mills sector. An implementation of this construction os also available in the package.

The main structure needed for the integration of the gradient flow equations is \texttt{FlowIntr}, where the details of the integration scheme are specified. These can be defined by functions such as \texttt{wfl\_euler(Float64,eps,tol)}, where \texttt{eps} is the stepsize and \texttt{tol} is the tolerance for the adaptive stepsize. This function defines the parameters for the Euler integration scheme with Wilson flow. The rest of the integration schemes can be defined by substituting ''\texttt{euler}'' for ''\texttt{rk2}'' or ''\texttt{rk3}'' for Runge-Kutta of order 2 and 3 respectively, and ''\texttt{wfl}'' for ''\texttt{zfl}'' for Zeuthen flow \cite{Ramos:2015baa}.

The integration of the flow equations is implemented in the function \texttt{flw(U, int::FlowIntr, ns::Int64, eps, gp, lp, ymws),} where \texttt{ns} is the number of steps in the integration, meaning that the total integration length will be \texttt{t=ns$\times$eps}. One can ommit \texttt{eps}, and the value of \texttt{int.eps} will be used.

An implementation of the adaptive step size integration is also available via the \texttt{flw\_adapt(U, int::FlowIntr, tend, epsini, gp, lp, ymws),} method. The argument \texttt{tend} is the total distance of integration and \texttt{epsini} is the value of the initial step for the integration and can be ommited, in which case \texttt{int.eps} will be used. In these methods, the integration step is recomputed every ten steps to keep the error of the integration below \texttt{int.tol}. In this case, the function will return two values: the number of steps in the integration and a vector with all the step sizes.

\subsection{Fermionic observables}

In this section we discuss the implementation of Wilson fermions in the package. The necessary parameters for fermionic meassurements are stored in the \texttt{DiracParam} structure

\begin{center}
  \texttt{dpar = DiracParam\{Float64\}(REP,m0,csw,theta,mu,cF)},
\end{center}where \texttt{REP} can take the values \texttt{SU2fund} and \texttt{SU3fund}, \texttt{m0} is the bare quark mass, \texttt{csw} is the Sheikholeslami-Wohlert coefficient \cite{Sheikholeslami:1985ij}, \texttt{theta} is a 4-vector with the phase-shift for the fermions in the boundaries\footnote{Each value of \texttt{theta} is not the phase in each direction, but the full phase factor, so $||\texttt{theta}_{i}||=1$ should be numerically imposed.}, \texttt{mu} is the twisted mass and \texttt{cF} is the improvement coefficient for the fermions with Dirichlet boundary conditions (also denoted as $\tilde{c_{t}}$) \cite{Luscher:1992an}.

The Dirac operator is implemented in the functions \texttt{Dw!}, \texttt{g5Dw!} and \texttt{DwdagDw!} with the syntax \texttt{Dw!(pso,U,psi,dpar::DiracParam,dws::DiracWorkspace,lp::SpaceParm)}, where \texttt{pso} and \texttt{psi} are the output and input quark fields respectively. Note that both fields have to be allocated before the call and the output field will be modified.

Another important function is the \texttt{Csw!(dws, U, gp, lp)} function. This updates the value of the field \texttt{dws.csw}, where the Sheikholeslami-Wohlert term is stored with the clover discretization. Once this is done, this term is added every time the Dirac operator is applied.

The Conjugate-Gradient (CG) solver is implemented and allows to compute correlation functions of fermionic fields. This can be done either by inverting the Dirac operator on a specific source via the \texttt{CG!} method or by using the \texttt{propagator!} function. This function has two methods: one utilizes normally distributed random noise sources, while the other inverts a source with a specific color, spin, and position. Similar functions to compute the boundary-to-bulk propagators for the Dirichlet boundary conditions are also available, allowing to compute different correlators such as $f_{P}$ and $f_{A}$ for SF boundary conditions.

An implementation of the Gradient flow for fermion fields following the lines of Ref.\cite{Luscher:2013cpa} is also available in the package. The functions described for the pure gauge cause are extended to include the integration of fermion fields, thanks to the multiple dispatch feature in Julia.

Unlike in the gauge scenario, estimating certain correlation functions leads to the integration of the adjoint flow equations. The numerical instabilities apearing for the gauge fields are overcome in a similar way as in Ref \cite{Luscher:2013cpa}. An integration with adaptive step size is performed first to fix a grid in the flow time. Then, \texttt{maxnsave} intermediate gauge fields in this grid are stored, and the fermion field is backflowed through the grid using the closest stored gauge field to reconstruct the necessary gauge field at each step. The function

\begin{center}
  \texttt{backflow(psi, U, Dt, maxnsave, gp, dpar, lp, int, ymws, dws),}
\end{center}implements this procedure, where \texttt{Dt} is the value of the final flow time, and \texttt{maxnsave} is the number of intermediate field configurations stored for the integration\footnote{Intermediate gauge configurations are stored in the CPU memory.}. The field \texttt{psi} is the fermion field at flow time \texttt{Dt}, while the gauge field \texttt{U} must be the gauge field at zero flow time. One important restriction is that \texttt{int} must be any order three Runge-Kutta, and if ommited the value \texttt{wfl\_rk3(Float64,0.01,1.0)} will be used. This function will modify the value of \texttt{psi} and return nothing.

\subsection{Performance}

The package provides performance details of most methods wothout specific benchmarking in the main program using the \texttt{TimerOutputs.jl} package. We have also benchmarked the performance between two devices for different processes in a $L/a = 32$ lattice: 50 integration steps of fourth order Omelyan, Gradient flow integration to $t/a^{2}=10.0$ with stepsize $\epsilon = 0.01$, backflow of a fermion field from $t/a^{2}=10.0$ and 1000 iterations of the CG. In all these cases, we obtain the expected scaling compatible with the devices specifications.

\begin{table}[h!]
  \centering
  \begin{tabular}{c|cccc}
    & HMC & YMGF & Bfl & Dw \\  \hline
    A100 & 6.6s & 80.0s & 92.7s & 13.9s\\ 
    H100 & 2.8s & 37.2s & 46.9s & 6.46s \\ 
    A/H & 2.34 & 2.15 & 1.97 & 2.15 \\
  \end{tabular}
\end{table}

We also compared the performance of the Nvidia A100 against CPUs of the Intel Xeon IvyBridge generation, obtaining an estimated performance of 1 GPU $\sim$ 80-100 CPUs.

\subsection{Input/Output}

Reading and writting both gauge configurations and fermion propagators is supported with native formats via the functions \texttt{save\_cnfg}, \texttt{read\_cnfg}, \texttt{save\_prop} and \texttt{read\_prod}. Other standard formats are also supported for reading gauge configurations such as the CERN format used in \cite{oqcd}.

\section{Some examples}

In this section, we will provide some examples to illustrate the usability of the package in some simple scenarios. We start by defining the main data structures with the parameters of the simulation and allocating the necessary fields and workspaces.

\begin{lstlisting}[style=custom]
using LatticeGPU
lp = SpaceParm{4}((16,16,16,16),(4,4,4,4),BC_PERIODIC,(0,0,0,0,0,0))
intsch = omf4(Float64,0.01,50)
# Gauge sector
U = vector_field(SU3{Float64},lp);
ymws = YMworkspace(SU3,Float64,lp);
gp = GaugeParm{Float64}(SU3{Float64},6.0,5/3,                            (1.0,1.0),(0.0,0.0),lp.iL)
# Quark sector
psi = scalar_field(Spinor{4,SU3fund{Float64}},lp);
dws = DiracWorkspace(SU3fund,Float64,lp);
dpar = DiracParam{Float64}(SU3fund,0.2,1.0,
                        (1.0,1.0,1.0,1.0),0.0,1.0)
\end{lstlisting}

Now we can start the simulation with a cold configuration and start generating our Monte Carlo chain, in this case of one hundred configurations. We also display the value of the plaquette and compute the quark propagator, with wich one could then compute any contraction.

\begin{lstlisting}[style=custom]
fill!(U,one(SU3{Float64}));
for i in 1:100
    dh,acc = HMC!(U,intsch,lp,gp,ymws)
    println("## Monte Carlo step ",i)
    println("## Plaquette : ",plaquette(U, lp, gp, ymws))
    Csw!(dws,U,gp,lp)
    niter = propagator!(psi, U, dpar, dws, lp, 10000, 1.0e-13, 1)
    println("## Inversion converged after ",niter," iterations.")
end
\end{lstlisting}
\section{Conclusions}

In this work we presented \texttt{LatticeGPU.jl}, an open-source package for Lattice simulations on GPUs developed in Julia. The power and flexibility of Julia combined with the high efficiency of GPUs provide a great enviroment for lattice simulations. It allows the simulation package to have competitive performance while also being easily extensible, both in expanding the current capabilities and in developing new functionalities.

The code version is robust, as several results from the literature and other codes have been reproduced. Future developments in the package can be expected in different directions, from the implementation of dynamical fermions and more sophisticated solvers to the implementation of less conventional tools such as the already implemented fermion flow.

The main advantage this code can present is a good ratio of efficiency and simplicity. The use of GPUs allows for a very high performance, and the use of Julia together with the fact that a front-end use of the package does not require specific GPU knowledge, allows for this good balance.

\acknowledgments

This work is partially supported by the Spanish Research Agency (Agencia Estatal de Investigación) through the grants IFT Centro de Excelencia Severo Ochoa CEX2020-001007-S and PID2021-127526NB-I00, funded by MCIN/AEI/10.13039/501100011033.

AR acknowledge support from the Generalitat
Valenciana grant CIDEGENT/2019/040, the European projects
H2020-MSCA-ITN-2019//860881-HIDDeN and 101086085-ASYMMETRY, and the national
projects CNS2022-136005 and PID2020-113644GB-I00.

We also acknowledge partial support from the project
H2020-MSCAITN2018-813942 (EuroPLEx) and the EU Horizon 2020 research
and innovation programme, STRONG2020 project, under grant agreement
No. 824093. Numerical calculations have been performed on the Hydra
and Ciclope installations at IFT, and local SOM clusters, 
funded by the MCIU with funding from the European Union NextGenerationEU
(PRTR-C17.I01) and Generalitat Valenciana, ASFAE/2022/020, 
Artemisa, funded by the European Union ERDF and Comunitat
Valenciana. We also thank A. Rago for access to the DeiC installation
at SDU for test purposes. F.P.P. and C.P. thank P. Fritzsch and
S. Sint for their kind hospitality and fruitful discussions.

\bibliographystyle{JHEP}
\bibliography{biblio}

\end{document}